# Bifunctional Luneburg-fisheye Lens Based on the Manipulation of Spoof Surface Plasmons

Jin Zhao, Yi-Dong Wang, Li-Zheng Yin, Feng-Yuan Han, Tie-Jun Huang, Pu-Kun Liu[a]

*State Key Laboratory of Advanced Optical Communication Systems and Networks, Department of Electronics, Peking University, Beijing,100084, China.*

[a)]Email: pkliu@pku.edu.cn

## ABSTRACT

Manipulation of spoof surface plasmons (SSPs) has recently intrigued enormous interest due to the capability of guiding waves with subwavelength footsteps. However, most of the previous studies, manifested for a single functionality, are not suitable for multifunctional integrated devices. Herein, a bifunctional Luneburg-fisheye lens is proposed based on a two-dimension metal pillar array. Firstly, by tuning the geometric dimension of the metal pillars in the array, its ability to precisely manipulate the excited SSPs along one direction is confirmed, achieving subwavelength focusing and imaging with the resolution up to $0.14\lambda$. Then, separately controlling the propagation of the SSPs along the orthotropic directions is further implemented, and the bifunctional Luneburg-fisheye lens is realized. The bifunctional lens is characterized as a Luneburg one along the *x*-axis, whereas in the *y*-axis, it presents the properties of a Maxwell fisheye lens. The experimental results almost immaculately match with the simulation ones. This bifunctional lens can validly reduce the system complexity and exert flexibility in multifunctional applications, while the proposed metal pillar-based design method broadens the application range of gradient refractive-index lens in the microwaves, terahertz, and even optical ranges.

Surface plasmon polaritons (SPPs) are produced by collective oscillations of electron gases near metallic surfaces and interfaces, which behave as surface waves in the optical regime.[1] The SPPs propagate along the interface between the metal and dielectric, simultaneously, decay exponentially along the direction perpendicular to the interface.[2] They have inherent features such as strong confinement and near-field enhancement effect. Attributing to the intriguing properties of SPPs, they have been applied to a wealth of designs including miniaturized photonic circuits,[3] super-resolution imaging,[4,5] high-sensitive sensors,[6,7] and detection.[8,9]

From the perspective of terahertz and microwave frequencies, however, the metals behave like the perfect electric conductor, along which the natural SPPs are severely weakened.[10] To achieve the analogous performance in microwave and terahertz regions, Pendry *et.al.* constructed the periodic hole array on a metal surface to mimic the optical SPPs properties, which are called spoof surface plasmons (SSPs).[11] Based on this concept, researchers endow SSPs with immense potentials, one of which is manipulating the propagation of SSPs. A series of novel devices have been corroborated on SSPs manipulation, including plasmonic waveguides,[12,13,14] super focusing of waves,[15,16] and sensing.[17]

In order to improve the integration and reduce the weight of the system, it is often necessary to introduce multiple functions into a single device. However, the aforementioned researches are less concerned about multifunctional devices, which are urgently required for multipurpose applications. Encouragingly, recent researches gradually pay more attention to this issue and propose diverse multifunctional devices such as the multichannel metasurface,[18,19] the reconfigurable multifunctional meta-grating,[20] and the bifunctional lens.[21,22,23] The bifunctional lens can fuse two functions in a single device, which availably reduces the systematic complexity. For example, the Luneburg lens and Maxwell fisheye lens are typical gradient-refractive-index (GRIN) devices. The former can transform an incident plane wave into a point source on the surface,[24] while the latter can refocus the point source on the surface from one side to the opposite side.[25] Both of them are widely utilized in the microwave, terahertz, and optical regions for information processing and communication applications.[26,27] If the Luneburg lens is integrated with the Maxwell fisheye lens in a single device, such as a waveguide crossing,[27] the flexibility may be strikingly increased.

In this paper, the metal pillars array is yielded to construct a bifunctional Luneburg-fisheye lens. The SSPs are not only capable to propagate on the surface of the metal pillars array, but also obtain adjustable effective refractive-index by modulating the size of the pillar. Based on the dispersion curves of the metal pillars with various sizes, gradient effective refractive-index (GERIN) of SSPs are obtained. A GRIN self-focus lens is firstly designed to verify the array's capability of possessing SSPs along one direction. The simulation results illustrate that the subwavelength focusing and imaging are fulfilled, indicating the feasibility of this design method. Further, the independent manipulation of SSPs along two orthogonal directions is achieved, based on which the bifunctional lens is implemented. The experiment results manifest that the Luneburg lens is realized along the $x$-axis while the Maxwell fisheye lens is presented along the $y$-axis. The bifunctional lens may be potential for the waveguide crossing and the SSPs-based design method can be facile applied to other devices, such as the bifunctional Eaton-fisheye lens, Eaton-Luneburg lens, and other GRIN lenses.

In our design, as illustrated in Fig. 1(a), the metal pillar on a flat metal base is employed as the unit to manipulate the propagation of SPPs. Assuming the unit is infinitely arranged in the $x$- and $y$-direction, the surface modes in this 2D array are derived.[28] To be simplified, the lattice constant, width, and depth are set as $a = a_x = a_y = 50$ μm, $w = w_x = w_y = 10$ μm, and $h = 150$ μm, respectively. Without loss of generality, we assume $k_y = 0$, and only the $x$-direction propagation modes are considered. For the TM-polarized surface modes, the dispersion equation[28] is derived as follows:

$$\cot(k_0 h) = \sum_{n=-\infty}^{n=+\infty} \frac{k_0 S_n^+ S_n^-}{\sqrt{k_{x,n}^2 - k_0^2}} \tag{1}$$

where $S_n^+ = \sqrt{w/a}\{\text{sinc}(k_{x,n}w/2) + \text{sinc}(k_{x,n}w/2) - (1 - w/a)\text{sinc}[k_{x,n}(a-w)/2]\}$, $S_n^- = \sqrt{w/a}\ \text{sinc}(k_{x,n}w/2)$, and $k_{x,n} = k_x + 2\pi n/a$. $n$ represents the harmonic order. It is worth noting that there are additional TE-like hybrid surface modes (hereinafter referred to as TE modes) in the structure, whose dispersion relation is demonstrated as:

$$\cot\left(K\frac{d}{2}\right) = \sum_{n=-\infty}^{n=+\infty} \frac{wK}{a\alpha_n} \coth\left(\alpha_n \frac{w}{2}\right) \text{sinc}^2\left(\frac{k_{x,n}w}{2}\right) \tag{2}$$

where $K = (k_0^2 - k_{c,m}^2)^{1/2}$, $k_{c,m} = (2m - 1)\pi/2h$ with $m$ being a positive integer representing the TE mode order, $k_{x,n} = k_x + 2\pi n/a$, $\alpha_n = (k_{x,n}^2 - K^2)^{1/2}$ and $d = a - w$. The dispersion relations are depicted in Fig. 1(b), which illustrate the simulation results fit well with the theoretical values. Note that the $TE_1$ mode curve (the lowest-order TE mode) is above the asymptotic frequency of $TM_0$ mode, that is, all of the TE modes are higher than the $TM_0$ model. Consequently, by restricting the working frequency, only the $TM_0$ mode is obtained in our design.

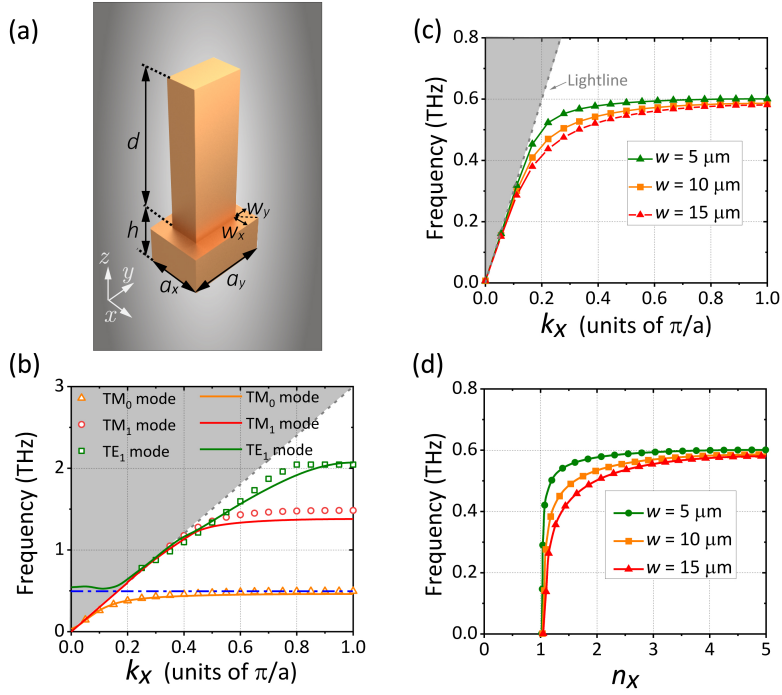

Figure 1. (a) Unit form. (b) Simulated (solid line) and the theoretical (scatters) dispersion relations of $TM_0$, $TM_1$, and $TE_1$-mode. The dash-dot line represents the asymptotic frequency of $TM_0$ mode. (c), (d) Dispersion relations and the effective refractive-index of SSPs with different $w$ ($a = 50$ μm, $d = 115$ μm).

As depicted in Fig. 1(c), the $w$ is changed to get multiple deflected dispersion curves. The $k_x$ represents the propagation constant of the SSPs traveling along the *x*-axis. The equivalent refractive-index of SSPs is calculated as $n_x = k_x / k_0$, where $k_0$ is the free space propagation constant at the corresponding eigenfrequency. Since $a_x = a_y$, $w_x = w_y$, the unit presents identical character along the *x*- and *y*- directions, which means $n_x = n_y$ at the same frequency. As illustrated in Fig. 1(d), at one certain frequency, a series of tailored $n$ can be obtained by adjusting the pillar size.

Before constructing a bifunctional lens that presents different GRIN features related to the orthogonal directions, we first need to figure out how to attain the GRIN along one single direction. Take the self-focus lens[29] as an example, it complies with the hyperbolic secant refractive-index distribution along one direction, i.e., $n(y) = n_0\text{sech}(\alpha y)$, where $n_0$ represents the refractive index at the central axis $y = 0$ and $\alpha$ is the gradient coefficient. Here, we set $n_0 = 3.9$ and $\alpha = 3400$.

To achieve this GRIN along the *y*-direction, the pillars' dimension is altered along the *y*-axis and keeps consistent along the *x*-axis. The elaborately designed lens and the effective refractive-index distribution are described in Fig. 2(a). All of the units have 10 different $w$ with $d = 115$ μm, $a = 50$ μm, and $h = 40$ μm. As plotted in Fig. 2(b), the incident wave propagates along the *x*-axis and converges to the central focus spot at the working frequency of 0.57 THz. Then, the wave diverges and gradually transforms into a plane SSPs wave. The FWHM of the spot is 0.14λ, which achieves the subwavelength focusing. Furthermore, as indicated in Fig. 2(c), three electric dipoles spaced 0.33λ apart with uniform phase are used as the sources. Their images are distinctly reconstructed on the imaging plane, which demonstrates the subwavelength imaging property of the lens. The above-mentioned traits indicate that the lens is successfully acquired and the design method of the GRIN lens along the single direction is available.

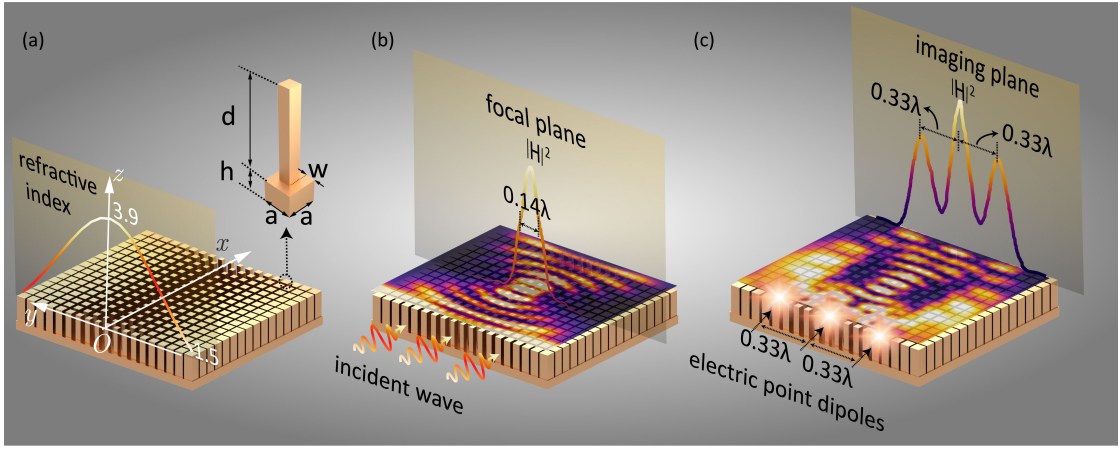

Figure 2. (a) Designed self-focus lens with the hyperbolic secant refractive-index distribution and the unit cell. (b) Characteristic of subwavelength focusing. (c) Feature of subwavelength imaging with three-point sources.

Next, we aim to establish the bifunctional lens with Luneburg and Maxwell fisheye characteristics along the *x*- and *y*-axis, respectively. Both the Luneburg lens and Maxwell fisheye lens are GRIN lens. Assuming $n_l$ represents the refractive index of the Luneburg lens, *R* denotes the lens radius and *r* is the radiation distance from the lens center. As depicted in Fig. 3(a), the refractive index monotonously increases as *r* decreases. The *priori* refractive-index distribution of the Luneburg lens is expressed as $n_l = [2 - (r/R)^2]^{1/2}$, with which it can focus an incident plane wave into a point source on the surface, as illustrated by the ray trace in Fig. 3(a). Analogously, for the Maxwell fisheye lens, the refractive-index $n_f$ follows the formula that $n_f = n_0 / [1 + (r/R)^2]$, where $n_0$ is the maximum refractive-index in the lens center ($n_0 = 2$ in this work). It has the capability to refocus a point source from the lens surface to another side of the lens.

To be rigorous, $n_x$ and $n_y$ cannot be immediately equal to $n_l$ and $n_f$, respectively. To gain better insight into this issue, the ray trace sketch is delineated in Fig. 3(b). As the ray traces gradually deflect from the optical axis (i.e., *x*- or *y*-axis), the angle between the ray's tangential direction and the optical axis becomes larger. For instance, at point A in the upper right inset of Fig. 3(b), the requested $n_l$ for the Luneburg lens is actually along the *x'*-direction, not the *x*-direction anymore.

To ascertain the relationship between $n_l$, $n_f$, $n_x$, and $n_y$, we take point A as an example. Assuming the angle between the *x*-axis and the ray's tangential direction at point A is $\theta_l$. Then, we rotate the *x-y* axes counterclockwise by $\theta_l$ to obtain another new *x'-y'* axes. The relationship between *x'-y'* axes and *x-y* axes is derived as:

$$\begin{cases} x' = x\cos\theta_l + y\sin\theta_l \\ y' = y\cos\theta_l - x\sin\theta_l \end{cases} \quad (3)$$

The refractive-index in the *x'-y'* coordinate system is explicitly expressed as follows:[22]

$$\begin{cases} n'_{xx} = n_x \cos^2\theta_l + n_y \sin^2\theta_l \\ n'_{xy} = n'_{yx} = (n_y - n_x)\cos\theta_l \sin\theta_l \\ n'_{yy} = n_y \cos^2\theta_l + n_x \sin^2\theta_l \end{cases} \quad (4)$$

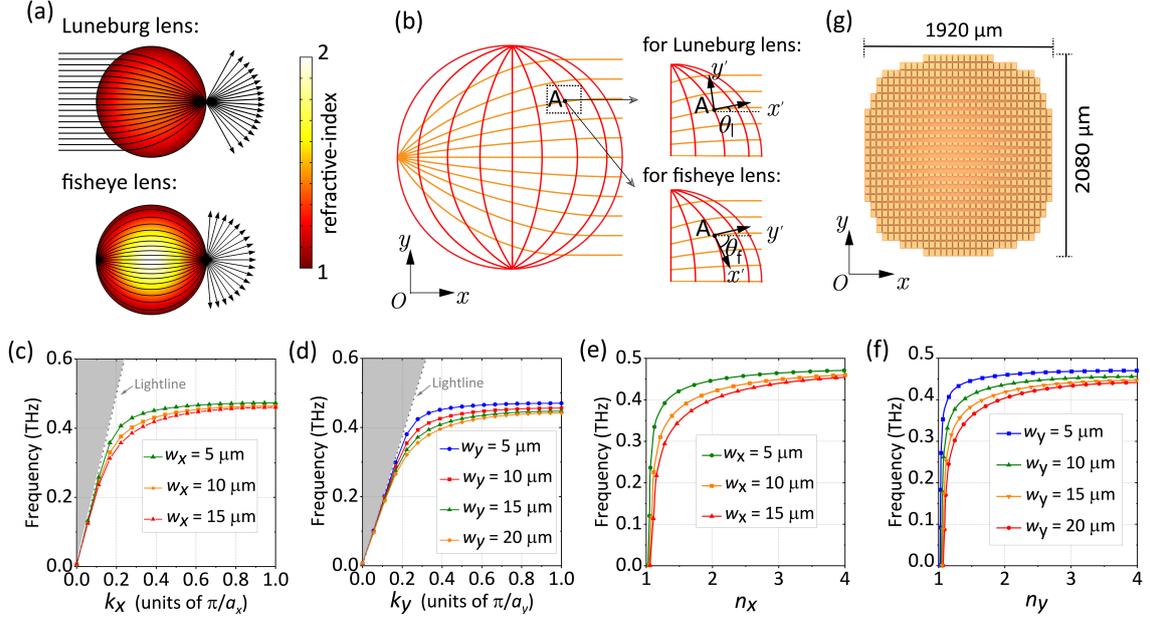

Figure 3. (a) Refractive-index distribution and ray trace (black line) of the Luneburg lens and Maxwell fisheye lens. (b) Overlapped ray trace of the Luneburg and Maxwell fisheye lens in the left view and transformation coordinate system for both the lens in the right view. (c) Dispersion curves of the unit under different $w_x$ ($w_y$ = 25 μm). (d) Dispersion curves under different $w_y$ ($w_x$ = 10 μm). (e) Effective refractive-index of SSPs along the $x$-direction under different $w_x$ ($w_y$ = 25 μm). (f) Effective refractive-index of SSPs along the $y$-direction under different $w_y$ ($w_x$ = 10 μm). (g) Schematic diagram of the designed bifunctional lens.

It is noticed that the cross-coupling item $n'_{xy}$ and $n'_{yx}$ can be negligible when $n_x$ and $n_y$ have slightly different. $n'_{xx}$ represents the effective refractive-index at point A along the $x'$-direction, which is also the required direction for refractive-index $n_l$. Thus, the transformational relation between $n_x$, $n_y$, and $n_l$ is shown as:

$$n_x \cos^2 \theta_l + n_y \sin^2 \theta_l = n_l \qquad (5)$$

Similarly, the Maxwell fisheye lens can be analyzed by repeating the aforementioned method. In the lower right inset in Fig. 3(b), the axes are rotated clockwise by the angle $\theta_f$, which is the included angle between the tangential ray trace at point A and the $x$-axis. This time the Equation (5) is replaced as:

$$n_x \cos^2 \theta_f + n_y \sin^2 \theta_f = n_f \qquad (6)$$

Hereto, accurate $n_x$ and $n_y$ at each point can be calculated with Equations (5) and (6) since $n_l$, $n_f$, $\theta_l$, and $\theta_f$ are already known. What we need to do is to match the pillar size with the requested $n_x$ and $n_y$ at each discrete position (center of each unit). Since the required $n_x$ and $n_y$ are mostly different in the same position, $w_x$ and $w_y$ of a unit cannot be equal anymore. With $a_x$ = 60 μm, $a_y$ = 80 μm, $d$ = 148.2 μm, and $h$ = 40 μm, the dispersion curves of the unit under different $w_x$ and $w_y$ are shown in Figs. 3(c) and (d), respectively. Then, the effective refractive-index curves are obtained in Figs. 3(e) and (f), suggesting that $n_x$, $n_y$ are capable to cover the required refractive-index ranges of Luneburg and Maxwell fisheye lens at the working frequency of 0.38THz. Consequently, the final unit dimension is $a_x$ = 60 μm, $a_y$ = 80 μm, $d$ = 148.2 μm, and $h$ = 40 μm with $w_x$ ranging from 5 to 10 μm and $w_y$ changing from 5 to 25 μm.

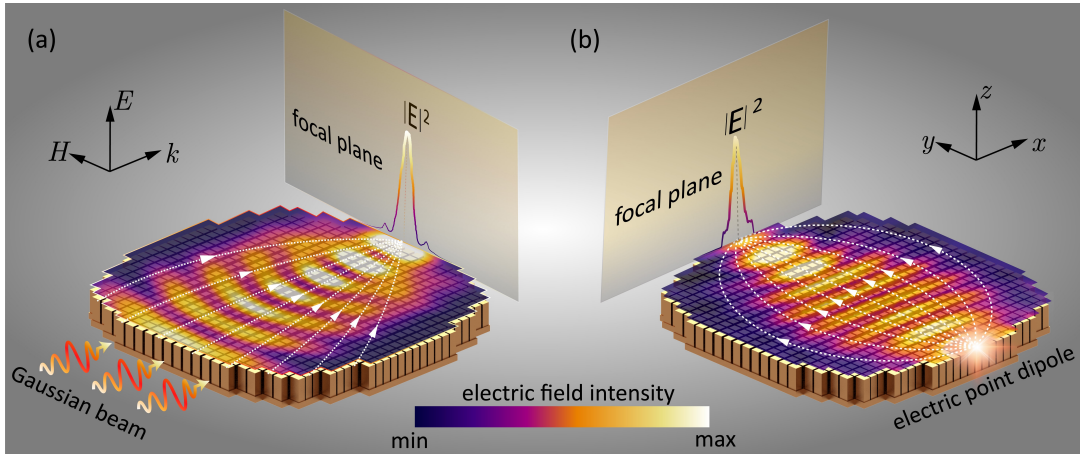

Figure 4. Schematic diagram of the Luneburg lens and Maxwell fisheye lens.

Through judicious design, the bifunctional lens containing 366 units is obtained. The sketch of the lens is shown in Fig. 3(g). The simulation is completed with commercial software COMSOL 5.4a. To verify the Luneburg lens property, a Gaussian beam propagating along the *x*-axis is employed to impinge on the lens. As shown in Fig. 4(a), with the electric field intensity converging to the opposite side, a focal spot is generated on the focal plane. For the testify of Maxwell fisheye lens, as depicted in Fig. 4(b), an electric point dipole placed 160 μm away from the lens is utilized as the excitation source. Expectedly, the electric field intensity is refocused on the focal plane.

Since the electric field of the Gaussian beam is polarized along the *z*-direction, we exhibit the 2-D electric field distribution of $E_z$ in Fig. 5(a). It also reveals that the lens focuses the Gaussian beam into a point source on the other side. Similarly, under the illumination of an electric point dipole polarized along the *z*-axis, the 2-D $E_z$ field for the Maxwell fisheye lens is demonstrated in Fig. 5(b). The electric field is refocused on the other side of the lens, which substantiates the function of a Maxwell fisheye lens.

With the consideration of our facilities, we experimentally verify the bifunctional lens in the microwave region. Actually, limited by the fabrications and testing conditions, experiments at the terahertz regime are relatively difficult. For the fabrication, our simulation parameters fall between the traditional machining and micromachining methods. On the one hand, the machining method will be time-consuming for our hundreds of pillars and the manufacture tolerance is hard to meet our parameter requirements. On the other hand, most of the micromachining methods allow the metal thickness from hundreds of nanometers to dozens of microns. Since our pillar height has surpassed this range, this method is dissatisfactory. For the test, the effective terahertz sources and detectors are scarce. Therefore, the experiment frequency is moved to 3.8 GHz and the lens dimension is amplified by 100 times, which is acceptable as the proposed method is valid from microwaves to terahertz regions.

The manufactural lens made of aluminium alloy is shown in Fig. 5(c), the whole length is 208 mm and the total width is 198 mm. Firstly, we demonstrate the Luneburg lens function, for which the incident wave is launched along the *x*-axis. The experimental schematic diagram is depicted in Fig. 5(d). A horn antenna is applied for the generation of the Gaussian beam, while an electric dipole is placed 2 mm above the lens surface for the detection of the 2-D $E_z$ field distribution in the white dotted box. The distance between the horn and the lens is 104 cm. The measured results in Fig. 5(e) indicate that the Gaussian beam is focused on the lens surface, which is almost perfectly consistent with the simulation.

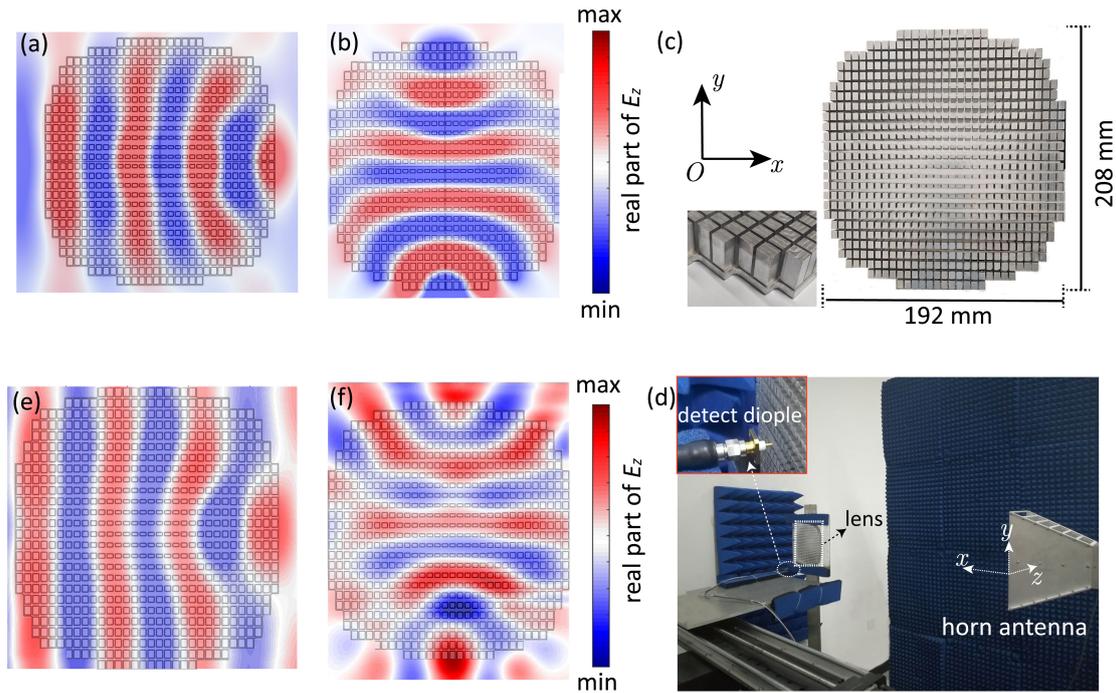

Figure 5. (a) Simulated real part of $E_z$ for the Luneburg lens. (b) Simulated real part of $E_z$ for the Maxwell fisheye lens. (c) Photograph of the manufactural lens and the zoomed-in view (lower left inset). (d) Experimental setup for the measurement of the field distribution for the Luneburg lens. (e) Measured real part of $E_z$ for the Luneburg lens. (f) Measured real part of $E_z$ for the Maxwell fisheye lens.

Then, another electric dipole identical to the detecting one is employed as the excitation source. It is placed 2 mm above the lens surface and 16 mm away from the lens side, which is consistent with the simulation scenarios [scene in Fig. 4(b)]. Likewise, the detecting dipole scans above the lens surface in the dotted box region and obtains the 2-D $E_z$ field, as shown in Fig. 5(f). The experiment results agree with the simulation, manifesting the Maxwell fisheye lens property.

At last, there remain some further discussions about this work. (1) The self-focus lens demonstrates the Fourier transform function, which means the image can be directly obtained without extra image retrieval processes. This simply constructed lens may provide a way in the terahertz real-time imaging field. (2) The proposed bifunctional lens combines the Luneburg with Maxwell fisheye lens. It may have the potential to be utilized as the waveguide crossing to reduce system complexity and improve application flexibility. (3) In general, the proposed methodology of designing a bifunctional GRIN lens is applicable for the other bifunctional lenses, for example, the bifunctional Eaton-Luneburg lens and the Eaton-Maxwell fisheye lens.[30,31] (4) It is noticed that the effective refractive-index of the metal pillar array can obtain a large range through decreasing the pillar's width, increasing the working frequency of the fundamental mode, or even raising the height of the pillar.

In summary, we propose a bifunctional lens based on the metal pillar array. The SSPs on the array obtain adjustable effective refractive-index by altering the pillars' size, resulting from which the propagation of SSPs can be manipulated. Based on this strategy, a self-focus lens is attained, indicating that the GREIN distribution along one direction can be realized. Then, we further introduce two different GREIN distributions along the orthogonal directions and construct a bifunctional Luneburg-fisheye lens. It behaves as a Luneburg lens along the *x*-direction and a Maxwell fisheye lens along the *y*-direction. The experimental results are consistent with the simulation ones, which demonstrates the feasibility of our scenario. The bifunctional lens may provide potential applications in the SSPs based waveguide crossing and the design methodology can provide a way to construct the GRIN lens in the microwaves, terahertz, and even optical ranges.

## ACKNOWLEDGMENTS

The authors would like to thank Dr. Shen Zheng of Aerospace Information Research Institute, Chinese Academy of Sciences for his help in the experiment. This work is sponsored by the National Key Research and Development Program under Grant No. 2019YFA0210203; National Natural Science Foundation of China under Grant No.61971013 and 61531002.

## DATA AVAILABILITY

The data that support the findings of this study are available from the corresponding author upon reasonable request.